%% file: main.tex
\documentclass[sigconf,natbib=false]{acmart}
\usepackage{algorithm}
\usepackage{algpseudocode}
\usepackage{graphicx}
\usepackage{textcomp}
\usepackage{xcolor}
\usepackage{tikz}
\usepackage{pgfplots}
\usepgfplotslibrary{groupplots}
\usepackage{pgfmath}
\pgfplotsset{compat=1.18}
\usepackage{subcaption}
\usepackage{multirow}
\usepackage{listings}

\RequirePackage[
  datamodel=acmdatamodel,
  style=acmnumeric,
  ]{biblatex}

\bibliography{sample-base}
\begin{document}
\title{Experiences Readying Applications for Exascale}
\author{Paul T. Bauman}
\email{paul.bauman@amd.com}
\orcid{0000-0003-3513-8264}
\affiliation{\institution{Advanced Micro Devices Inc.}\city{Austin}\state{Texas}\country{USA}}
\author{Reuben D. Budiardja}
\email{reubendb@ornl.gov}
\orcid{0000-0003-0395-8532}
\affiliation{\institution{Oak Ridge National Laboratory}\city{Oak Ridge}\state{Tennessee}\country{USA}}
\author{Dmytro Bykov}
\email{bykovd@ornl.gov}
\orcid{0000-0002-6668-4586}
\affiliation{\institution{Oak Ridge National Laboratory}\city{Oak Ridge}\state{Tennessee}\country{USA}}
\author{Noel Chalmers}
\email{noel.chalmers@amd.com}
\orcid{0000-0002-1293-7525}
\affiliation{\institution{Advanced Micro Devices Inc.}\city{Austin}\state{Texas}\country{USA}}
\author{Jacqueline Chen}
\email{jhchen@sandia.gov}
\orcid{0000-0002-9268-0634}
\affiliation{\institution{Sandia National Laboratories}\city{Albuquerque}\state{New Mexico}\country{USA}}
\author{Nicholas Curtis}
\email{nicholas.curtis@amd.com}
\orcid{0000-0002-0303-4711}
\affiliation{\institution{Advanced Micro Devices Inc.}\city{Austin}\state{Texas}\country{USA}}
\author{Marc Day}
\email{marc.day@nrel.gov}
\orcid{0000-0002-1711-3963}
\affiliation{\institution{National Renewable Energy Laboratory}\city{Golden}\state{Colorado}\country{USA}}
\author{Markus Eisenbach}
\email{eisenbachm@ornl.gov}
\orcid{0000-0001-8805-8327}
\affiliation{\institution{Oak Ridge National Laboratory}\city{Oak Ridge}\state{Tennessee}\country{USA}}
\author{Lucas Esclapez}
\email{lucas.esclapez@nrel.gov}
\orcid{0000-0002-2438-7292}
\affiliation{\institution{National Renewable Energy Laboratory}\city{Golden}\state{Colorado}\country{USA}}
\author{Alessandro Fanfarillo}
\email{alessandro.fanfarillo@amd.com}
\orcid{0000-0003-3487-7452}
\affiliation{\institution{Advanced Micro Devices Inc.}\city{Austin}\state{Texas}\country{USA}}
\author{William Freitag}
\email{chip.freitag@amd.com}
\orcid{0009-0001-6051-1975}
\affiliation{\institution{Advanced Micro Devices Inc.}\city{Austin}\state{Texas}\country{USA}}
\author{Nicholas Frontiere}
\email{nfrontiere@anl.gov}
\orcid{0009-0005-8598-4292}
\affiliation{\institution{Argonne National Laboratory}\city{Lemont}\state{Illinois}\country{USA}}
\author{Antigoni Georgiadou}
\email{georgiadoua@ornl.gov}
\orcid{0000-0002-0977-6310}
\affiliation{\institution{Oak Ridge National Laboratory}\city{Oak Ridge}\state{Tennessee}\country{USA}}
\author{Joseph Glenski}
\email{glenski@hpe.com}
\orcid{0009-0009-5432-3773}
\affiliation{\institution{Hewlett Packard Enterprise}\city{Bloomington}\state{Minnesota}\country{USA}}
\author{Kalyana Gottiparthi}
\email{gottiparthik@ornl.gov}
\orcid{0000-0002-1354-0255}
\affiliation{\institution{Oak Ridge National Laboratory}\city{Oak Ridge}\state{Tennessee}\country{USA}}
\author{Marc T. Henry de Frahan}
\email{marc.henrydefrahan@nrel.gov}
\orcid{0000-0001-7742-1565}
\affiliation{\institution{National Renewable Energy Laboratory}\city{Golden}\state{Colorado}\country{USA}}
\author{Gustav R. Jansen}
\email{jansengr@ornl.gov}
\orcid{0000-0003-3558-0968}
\affiliation{\institution{Oak Ridge National Laboratory}\city{Oak Ridge}\state{Tennessee}\country{USA}}
\author{Wayne Joubert}
\email{joubert@ornl.gov}
\orcid{0000-0003-4771-998X}
\affiliation{\institution{Oak Ridge National Laboratory}\city{Oak Ridge}\state{Tennessee}\country{USA}}
\author{Justin G. Lietz}
\email{lietzjg@ornl.gov}
\orcid{0000-0002-8398-5524}
\affiliation{\institution{Oak Ridge National Laboratory}\city{Oak Ridge}\state{Tennessee}\country{USA}}
\author{Jakub Kurzak}
\email{jakub.kurzak@amd.com}
\orcid{0000-0002-9697-0145}
\affiliation{\institution{Advanced Micro Devices Inc.}\city{Austin}\state{Texas}\country{USA}}
\author{Nicholas Malaya}
\email{nicholas.malaya@amd.com}
\orcid{0000-0001-6259-7453}
\affiliation{\institution{Advanced Micro Devices Inc.}\city{Austin}\state{Texas}\country{USA}}
\author{Bronson Messer}
\email{bronson@ornl.gov}
\orcid{0000-0002-5358-5415}
\affiliation{\institution{Oak Ridge National Laboratory}\city{Oak Ridge}\state{Tennessee}\country{USA}}
\author{Damon McDougall}
\email{damon.mcdougall@amd.com}
\orcid{0009-0008-5865-9322}
\affiliation{\institution{Advanced Micro Devices Inc.}\city{Austin}\state{Texas}\country{USA}}
\author{Paul Mullowney}
\email{paul.mullowney@amd.com}
\orcid{0000-0002-1504-5178}
\affiliation{\institution{Advanced Micro Devices Inc.}\city{Austin}\state{Texas}\country{USA}}
\author{Stephen Nichols}
\email{nicholsss@ornl.gov}
\orcid{0000-0003-3484-2735}
\affiliation{\institution{Oak Ridge National Laboratory}\city{Oak Ridge}\state{Tennessee}\country{USA}}
\author{Matthew Norman}
\email{normanmr@ornl.gov}
\orcid{0000-0003-4764-3348}
\affiliation{\institution{Oak Ridge National Laboratory}\city{Oak Ridge}\state{Tennessee}\country{USA}}
\author{Thomas Papatheodore}
\email{papatheodore@ornl.gov}
\orcid{0000-0002-6991-4332}
\affiliation{\institution{Oak Ridge National Laboratory}\city{Oak Ridge}\state{Tennessee}\country{USA}}
\author{Jon Rood}
\email{jon.rood@nrel.gov}
\orcid{0000-0002-7513-3225}
\affiliation{\institution{National Renewable Energy Laboratory}\city{Golden}\state{Colorado}\country{USA}}
\author{Philip C. Roth}
\email{rothpc@ornl.gov}
\orcid{0000-0001-9583-1103}
\affiliation{\institution{Oak Ridge National Laboratory}\city{Oak Ridge}\state{Tennessee}\country{USA}}
\author{Sarat Sreepathi}
\email{sarat@ornl.gov}
\orcid{0000-0002-4978-9423}
\affiliation{\institution{Oak Ridge National Laboratory}\city{Oak Ridge}\state{Tennessee}\country{USA}}
\author{James White III}
\email{trey.white@hpe.com}
\orcid{0009-0005-2186-075X}
\affiliation{\institution{Hewlett Packard Enterprise}\city{Bloomington}\state{Minnesota}\country{USA}}
\author{Noah Wolfe}
\email{noah.wolfe@amd.com}
\orcid{0000-0002-7935-421X}
\affiliation{\institution{Advanced Micro Devices Inc.}\city{Austin}\state{Texas}\country{USA}}


\acmConference[SC'23]{a}{Nov. 12--17, 2023}{Denver, CO, USA}
\begin{abstract}
  The advent of Exascale computing invites an assessment of existing best practices for developing application readiness on the world’s largest supercomputers. This work details observations from the last four years in preparing scientific applications to run on the Oak Ridge Leadership Computing Facility's (OLCF) Frontier system. This paper addresses a range of topics in software including programmability, tuning, and portability considerations that are key to moving applications from existing systems to future installations. A set of representative workloads provides case studies for general system and software testing. We evaluate the use of early access systems for development across several generations of hardware. Finally, we discuss how best practices were identified and disseminated to the community through a wide range of activities including user-guides and trainings. We conclude with recommendations for ensuring application readiness on future leadership computing systems. 
\end{abstract}


\maketitle

\renewcommand{\shortauthors}{Malaya et al.}

\section{Introduction}

On May 7th, 2019, the Oak Ridge Leadership Computing Facility (OLCF) announced a contract with Cray (later acquired by HPE) and AMD to build the Frontier supercomputer at Oak Ridge National Laboratory (ORNL). This machine was intended not only to exceed an exaflop of sustained double precision compute as measured by High-Performance Linpack (HPL), but also to provide significantly enhanced performance in scientific applications across a wide range of domains when the machine entered production. However, at that time few, if any, applications were ready to leverage the computational capabilities of an exascale machine. The unique architectural system design necessary for Frontier to efficiently achieve exascale had not yet been propagated to software codebases. 
Thus, the deployment of an exascale computer, itself a considerable undertaking, also necessitated a focused software effort for readying scientific applications to avail themselves of the full capabilities of the machine. 

Application readiness is a widely recognized challenge in the high performance computing (HPC) community, necessitating close partnerships between the computing facility, system integrator, and hardware vendors to ensure preparation in time for system deployment\cite{vazhkudai2018design,10.1007/978-3-031-10419-0_6,8024140,8024138,8960361}. However, while the essential challenge is not unique, the magnitude of the task has only increased with the recent growth in heterogeneous hardware, multiple programming models, and the unprecedented levels of parallelism required to effectively use an exascale computer. 

To address these challenges, a Frontier~\textit{Center of Excellence} (COE) was formed with the expressed purpose of pooling key staff from across HPE, AMD, and ORNL to act as a central repository of knowledge and expertise around application readiness and optimization. The COE acts as a focal point of application ``co-design'', coordinating efforts across a range of competencies including hardware, software, algorithms, and computational science. The Frontier COE was also intended to rapidly disseminate best practices to users and other leadership supercomputing centers. 

The purpose of this document is to detail the observations from the COE over the last four years of work readying scientific applications for the 
OLCF Frontier system and provide a careful assessment of current best-practices, along with novel methods, for application porting and tuning on the world's largest supercomputers. 

The rest of the paper is organized as follows. Software testing and readiness are discussed in Section~\ref{software}.
Next, Section~\ref{applications} discusses several examples of select applications across a representative range of scientific domains and computational motifs. Section~\ref{early-hw} briefly describes access to precursor hardware platforms. Section~\ref{training} then documents how the
lessons from individual application teams were rapidly disseminated to the broader user-base. Finally, the paper ends with Section~\ref{lessonslearned}, which discusses novel lessons learned in the course of the project, along with recommendations for other HPC centers focused on application readiness.  


\section{Software Testing and Readiness }
\label{software}

At the start of the project in 2019, few applications were ready to leverage the computational capabilities of an exascale machine. 
It was recognized that porting and optimizing codes for Frontier would be a considerable effort. To lessen the burden on application teams, Frontier's software was targeted at \textit{performance portability}, with the aim of enabling applications to run efficiently on Frontier and on other systems with different node designs and hardware vendors. 
%
The programming strategies discussed subsequently are AMD's HIP and OpenMP\textregistered  via target offload. Applications using abstraction frameworks such as Kokkos or RAJA to achieve performance portability are discussed in Sections \ref{E3SM}, \ref{pele}, and \ref{LAMMPS}.

\subsection{Early Evaluation of AMD's HIP}
\begin{figure*}
    \centering
    \includegraphics[width=0.8\linewidth,height=0.45\linewidth]{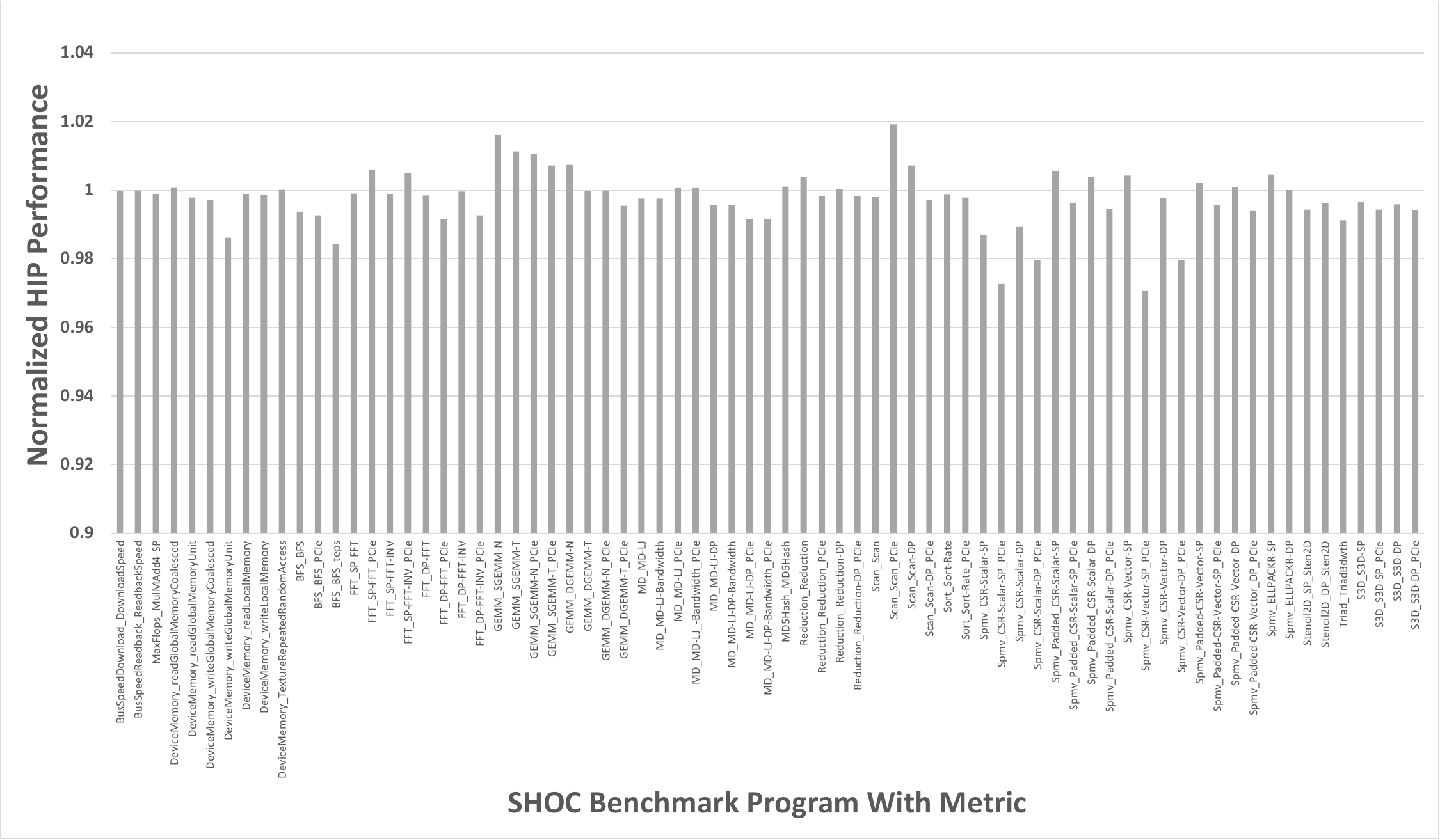}
    \caption{Performance of HIP on the SHOC benchmarks relative to CUDA versions running on OLCF Summit.  Note the Y-axis range is from 0.9 to 1.05.  Average normalized HIP performance was 99.8\% of CUDA performance when considering data transfer costs, 99.9\% without. In general, the overhead of using HIP vs. CUDA across a range of computational kernels is minimal.}
    \label{fig:hip-shoc}
\end{figure*}

HIP (Heterogeneous-compute Interface for Portability) is a portability layer that enables code to run on GPUs produced
by multiple vendors, with initial support for AMD and NVIDIA
GPUs.
To ease the porting of existing code to AMD GPUs, HIP's programming
interface was designed to be similar to recent (but not necessarily
the latest) versions of NVIDIA's CUDA API. AMD's HIP implementation
provided a ``\texttt{hipify}'' tool to produce HIP code from CUDA code.
As an early, partial evaluation of HIP's functionality and performance,
OLCF personnel used AMD's \texttt{hipify} tool to convert the CUDA implementations 
of the SHOC benchmark programs\cite{shoc1} to HIP and compared the performance of
both versions when run on OLCF's Summit system with its NVIDIA GPUs.
In most cases, the \texttt{hipify} tool converted the bulk of the code 
automatically, with the primary exception being code that used outdated
CUDA syntax. 
As shown in Figure~\ref{fig:hip-shoc}, the performance of the HIP
implementations was similar to that of the CUDA versions.
These findings are as expected for two reasons.
First, because the HIP implementation is a header-only
library when targeting NVIDIA GPUs, the HIP SHOC programs are CUDA executables once
compiled and linked.
And second, most SHOC programs are designed to focus on a particular
computation or data access pattern and so involved a small number of
GPU kernels.
The findings suggested the HIP approach was a feasible
approach for targeting GPUs from multiple vendors with a single code base.

The similarity between CUDA and HIP has allowed an alternative strategy to converting codebases to HIP. The strategy uses a single header file with macros to convert between CUDA and HIP calls depending on the build environment~\cite{ChollaFrontier}. The application code may remain in CUDA and evolve using either CUDA or HIP, as long as the functionality exists in both APIs. 

Although the similarity of HIP to CUDA decreases the effort required to port existing CUDA code to target AMD GPUs, it can foster the incorrect assumption among developers that \emph{every} CUDA feature from the \emph{latest} CUDA version is, or will be, provided by HIP.  Careful and repeated messaging to developers is needed to clarify which CUDA/NVIDIA GPU features will be replicated to correctly set expectations and to direct effort toward finding alternative, more portable approaches with comparable performance.


\subsection{OpenMP Offloading}
Several applications on Frontier rely on OpenMP target offload for GPU acceleration.  Advantages of OpenMP offload include simplified long-term maintenance of a codebase while enabling portability across multiple GPU architectures. In general, OpenMP codes did not achieve performance parity to codes ported with HIP. There were, however, several common strategies observed to be useful for applications using OpenMP to achieve high performance.

In general, loops can be accelerated with familiar OpenMP syntax. However, developers should use a large, structured \texttt{TARGET DATA} region around key performance regions of the code. These large regions should use persistent data arrays and variables on the GPUs via \texttt{MAP} to avoid repeated data movement between host and device. \texttt{OMP\_TARGET\_ALLOC} allocates such persistent arrays on the GPUs. 

Within a \texttt{TARGET DATA} region, data can be synchronized between host and device via a \texttt{TARGET UPDATE TO}/\texttt{FROM}. This synchronization leverages \texttt{NOWAIT} clauses for concurrent execution between host and device. \texttt{DETACH} can be used to overlap independent operations, and investigations are underway to identify its best use in the application codes such as GESTS~\cite{gests_sc19}. A \texttt{USE\_DEVICE\_PTR} can be used within a \texttt{TARGET DATA} region to instruct the compiler to use the GPU pointer of the array in function calls and to enable GPU-Aware MPI calls. Finally, unstructured \texttt{TARGET DATA ENTER}/\texttt{EXIT} pairs move data to/from the GPU when it's best to exclude the data from a structured \texttt{TARGET DATA REGION}.

\section{Application Experiences}
\label{applications}
The purpose of this section is to discuss applications that have demonstrated a significant performance improvement moving to Frontier, and to capture and disseminate the lessons learned and best practices the code teams have observed running on OLCF's Frontier system at large system scale (\textit{e.g.}, across hundreds or thousands of nodes). 
The optimization strategies presented subsequently encompass a multitude of topics ranging from performance portability, library use, kernel launch latencies, kernel fusion, and the use of reduced precision data types. These motifs, and the application examples they map to, are summarized in Table~\ref{tab:motif}. 
The speed-ups each application observed moving from OLCF-5 (Summit) to OLCF-6 (Frontier) are listed in~ Table~\ref{tab:speed-up}.

\begin{table}[tbp]
\caption{Application Porting Motifs}
\begin{center}
\begin{tabular}{ c | c }
 Porting Motif & Applications \\  
 \hline \hline
CUDA/HIP Porting & GAMESS, CoMet, \\
& NuCCOR, Coast \\
\hline
Library Tuning & GAMESS, LSMS, GESTS, \\
& CoMet, LAMMPS \\
\hline
Performance Portability & GESTS, ExaSky, E3SM, \\
& NuCCOR, Pele \\
\hline
Kernel Fusion/Fission & E3SM, Pele, LAMMPS \\
\hline
Algorithmic Optimizations & LSMS, ExaSky, E3SM, \\
& CoMet, Pele, LAMMPS \\
\label{tab:motif}
\end{tabular}
\end{center}
\end{table}

The applications selected in this section generally come from either Center for Accelerated Application Readiness (CAAR) or Exascale Computing Project (ECP) Application Development (AD) and Software Technology (ST) portfolios. In general, both CAAR and ECP applications had access to early hardware platforms (Section~\ref{early-hw}) and dedicated expert support  (Section~\ref{training}) and represent a sophisticated group of users with existing GPU programming experience and readiness. 

\begin{table}[tbp]
\caption{Observed application speed-ups from OLCF-5 (Summit) to OLCF-6 (Frontier)}
\begin{center}
\begin{tabular}{ c c }
 Application & Measured Speed-up (Frontier/Summit) \\ 
 \hline
GAMESS & 5 \\
LSMS & 7.5 \\
GESTS & 5 \\ 
ExaSky & 4.2 \\  
CoMet & 5.2 \\ 
NuCCOR & 6.1 \\
Pele & 4.2 \\ 
COAST & 7.4 \\ 
\label{tab:speed-up}
\end{tabular}
\end{center}
\end{table}

\subsection{GAMESS}
The General Atomic and Molecular Electronic Structure System (GAMESS) project is an established ab initio quantum chemistry package developed and maintained by the members of the Gordon research group at Iowa State University~\cite{GAMESS}. 
The GAMESS program has a broad user base of 150,000 in >100 countries. The code is available to the users at no cost via web download. 


GAMESS is predominantly legacy Fortran code with newer C/C++ development parts. The new developments were added and maintained as libraries inside the main code: LibCChem, LibCInt, and later EXESS~\cite{doi:10.1021/acs.jpca.0c02249}.
GAMESS is designed as a portable computational chemistry software package with the ability to run on many platforms and architectures. This has required transitioning from serial execution to vector and parallel execution while maintaining the full coverage of capabilities. The main computational motif inside GAMESS is dense linear algebra operations. As such, GAMESS depends on BLAS and diagonalization libraries, and used MAGMA and ROCm\texttrademark  for linear algebra operations. The LiBCChem/EXESS library inside GAMESS depends on Global Arrays, EIGEN, and CUDA/HIP. Parallel execution is accomplished through the generalized distributed data interface (GDDI) which relies on MPI as its communication layer. To maximize code portability at the exascale, directive-based approaches, such as OpenACC and OpenMP, were also exploited.

From the perspective of exascale computing the GAMESS team decided to further develop fragmentation methods: Fragment Molecular Orbital and Effective Fragment Molecular Orbital methods (FMO and EFMO, respectively), where computations are performed on the level of a single fragment using newly developed libraries~\cite{10046114}. The libraries are object-oriented C++ codes developed for both CPU and GPU and are distributed with GAMESS. LibCChem/EXESS includes codes for Rys quadrature two-electron integrals, Hartree-Fock (HF), MP2 and CCSD(T) coupled cluster theory. The fragments in the FMO scheme can be computed independently, and hence an efficient linear-scaling algorithm can be formulated. 

The development in the LibCChem/EXESS libraries was primarily using CUDA, and the code was then ported to HIP. The combined EXESS+GAMESS version was created during an early hackathon at OLCF. Initial testing on the MI250X after HIPification on Crusher indicated kernels running at almost double the flop rate of the V100 GPUs on Summit. A number of key optimizations for the memory transfer were introduced and resulted in substantial improvement of the RI-MP2 code being able to run at nearly peak device performance. A direct comparison between the Summit V100 and the AMD MI250X was not performed across the entire application, but the CUDA/HIP RI-MP2 libraries within GAMESS were evaluated in part. A speedup of 5x was observed in the fragment-level HIP RI-MP2 code within LibCChem/EXESS libraries.

The code was also tested on Frontier using the Many Body Expansion Fragmentation method on 128 to 512 nodes for a system comprised of 935 water molecules. An example comprised of ~75k atoms of an ionic liquid model system used 1024 and 2048 nodes at a time. ROCm 5.4 was used in conjunction with MAGMA to include a more efficient divide and conquer implementation of symmetric eigen solver. The code has shown excellent performance and nearly ideal linear scaling up to 2K nodes of the system. The GAMESS team has received INCITE allocation time at OLCF and is now in process of performing scientific runs. 

\subsection{LSMS}
The Locally Self-consistent Multiple Scattering code (LSMS) \cite{Wang1995, Eisenbach2017b} is a first principles application for solving the Schr\"odinger equation of electrons within a solid using density functional theory. It achieves linear scaling in the total number of atoms in the system by using a real space implementation of multiple scattering theory to solve the Kohn-Sham equations of density functional theory. LSMS was developed at ORNL and used on previous OLCF computer architectures. It was one of of the codes selected for the OLCF CAAR in preparation for Frontier.

As the computation in LSMS for spherical atomic potentials (the current main production mode) is dominated by linear algebra operations on non-Hermitian double precision complex dense matrices, the first step in moving to a new architecture is to use the vendor provided linear algebra libraries. In particular LSMS requires matrix-matrix multiplications (such as BLAS ZGEMM) and linear solvers, usually via LU factorization (e.g. LAPACK's ZGETRF and ZGETRS).
The original implementation of the GPU kernels for previous systems (e.g. OLCF's Titan and Summit) employed CUDA and calls to the cuBLAS library \cite{Eisenbach2017}. To prepare for Frontier, the GPU implementation increased the use of dense linear solver available in libraries. Thus, we replaced the block inversion algorithm by the LU factorization routines available in rocSOLVER (i.e. \texttt{rocsolver\_zgetrf} and \texttt{rocsolver\_zgetrs}). While both approaches have $O(N^3)$ scaling with the LIZ-matrix size, and the \texttt{zblock\_lu} algorithm has a slightly lower total floating point operation count, we observe better performance for the direct solution of the LIZ $\tau$ matrices using the rocSOLVER routines.
These library calls are supplemented by HIP GPU kernels that transform the LSMS problem into a form suitable for the linear solver. 

The second set of operations that were ported to HIP are the kernels needed to construct the structure constants that encode the system geometry and to assemble the KKR matrix. As these operations do not map readily onto standard library routines, we chose a HIP implementation of these algorithms. By employing kernel profiling we were able to identify bottlenecks in the first implementation of these kernels related to integer index and address calculations that interfered with the floating point operations on the MI250X GPUs. Rearranging these operations achieved significantly improved performance. Comparing the performance of a calculation for FePt systems, we see a per-GPU performance improvement of $\approx$7.5x on Frontier MI250X GPUs compared to Summit's V100.

\subsection{GESTS} \label{gests_subsection}
The GPUs for Extreme-Scale Turbulence Simulations (GESTS) team is part of the CAAR project and investigates turbulent flows across a wide range of scales via a Pseudo-Spectral Direct Numerical Simulation (PSDNS) algorithm \cite{gests_sc19}. DNSs of turbulence are well-suited to leadership computing as they use both the compute capabilities and the memory capacity offered by Exascale systems such as Frontier to probe high Reynolds number conditions\cite{leadershipdns,lee2013petascale}. 

The GESTS codes are written in Fortran 95 around a custom-built 3D FFT algorithm. The initial GPU-enabled version of the code-base relied solely upon CUDA functionality for data management, asynchronous operations, and the FFT computations. When moving to AMD hardware, vendor-specific functionality was limited to the core FFT functions, and OpenMP offloading was used to manage data movement between the host and device, to enable GPU-Direct MPI communications, and to accelerate a variety of array operations on the GPUs. 

Two variations of the PSDNS algorithm were developed: a \textit{Slabs} 1D- and a \textit{Pencils} 2D-domain decomposition. The \textit{Slabs} version is more efficient because it requires one fewer MPI communication cycle during both the forward and inverse FFT transforms than the \textit{Pencils} version.  However, for an $N^3$ problem, the \textit{Slabs} version is limited to $N$ MPI ranks, while the \textit{Pencils} version has a greater upper limit of $N^2$ MPI ranks. This extra flexibility makes the \textit{Pencils} code appropriate for large problems when memory-per-node becomes restrictive. 

As part of CAAR, GESTS targeted a $4\times$ improvement in a project-specific Figure of Merit (FOM). GESTS defined its FOM as $N^3/t_{\text{wall}}$ where $N^3$ is the number of grid points and $t_{\text{wall}}$ is the average runtime per time step.  The reference FOM was chosen as $N^3=18,432^3$, which was the largest problem completed during an INCITE 2019 project on Summit \cite{gests_sc19} using a highly performant, CUDA-based PSDNS algorithm.  Both versions of the ported code demonstrated an improvement of the FOM in excess of $5 \times$ on 4096 Frontier nodes using 32,768 MPI ranks for the $N^3=32,768^3$ problem. 

\subsection{ExaSky}
The ExaSky ECP project employs the HACC (Hardware/Hybrid Accelerated Cosmology Code) code ~\cite{frontiere2022farpoint, frontiere2022simulating, habib2017}, a particle-based cosmology framework. ORNL and the Science Engagement section have been contributors to the exascale preparation of the HACC code. Other collaborators are the Argonne National Laboratory, Los Alamos National Laboratory, and Lawrence Berkeley National Laboratory.

The ExaSky project intends to use Frontier with HACC to address multiple-level cosmological problems of interest ~\cite{heitmann2021}. To achieve this aim, a series of simulations with various volumes will be created to test initial value problems that best suit the scientific goals.

The HACC code has been under preparation to produce cosmological simulations of different implementations: (1) large-volume, high-mass resolution gravity-only simulations, (2) large-volume, high-mass resolution hydrodynamic simulations, and (3) small-volume, high-mass resolution hydrodynamic simulations. The code is a hybrid MPI-OpenMP implementation and only depends on an external FFT library.

Running HACC on the early access systems Poplar and Tulip (detailed in Section \ref{early-hw}) identified a challenge in using both HIP and OpenMP together. In this case, HACC used HIP to target execution on the GPU, and OpenMP for both asynchronous computing on the CPU and launching HIP kernels to the GPU. However, early compiler offerings didn't offer full support for both HIP and OpenMP in the same compilation unit. 
Developing general guidelines for building with both HIP and OpenMP on COE machines was a co-design effort across the code team, hardware vendor, and system integrator. 

Early performance assessments of the HIP implementation showed improved performance on second generation early access systems using AMD’s MI100s when compared to the NVIDIA Volta V100s available in OLCF Summit. Only one gravity kernel of the six of interest showed worse performance when using the AMD nodes. 
This change in performance for the gravity part of the code was connected to the use of the wavefront number size of 64 (the native wavefront size of the AMD GPU architecture) instead of 32 (the native wavefront size on the NVIDIA GPU architecture). Nevertheless, all major kernels demonstrated successful use of the Crusher system and had speed-ups compared to the Spock and Summit machines (see Section~\ref{early-hw} for details on the early access systems). 

The Frontier target at 8,192 nodes (32,768 GPUs) was a weak scaling benchmark that aimed to increase the Summit FOM by a factor of four; the measured speedup was found to be 4.2x. Successful runs on Frontier indicated the code performed well (gravity-only and hydro) and achieved a FOM of about 230x with respect to the original full machine baseline measured on the Theta supercomputer at Argonne National Laboratory. 

%

\subsection{E3SM} \label{E3SM}
The Energy Exascale Earth System Model---Multiscale Modeling Framework (E3SM-MMF) \cite{caldwell2019doe,norman2022unprecedented} is an ECP project with a throughput target of 1,000-2,000$\times$ realtime simulation. To achieve this ambitious goal at high spatial resolution, significant strong scaling is needed, which increases relative parallel overheads and decreases the per-node workload available to GPU accelerators. This means E3SM-MMF is highly sensitive to latency, and particularly allocations, deallocations, and kernel launches. 

Several strategies were used by E3SM to achieve performance optimizations and minimize latencies. Kernels that have little code were merged into larger kernels, as long as global algorithmic dependencies did not prohibit the fusion. This reduced the relative cost of kernel launch overheads. Conversely, some kernels with large amounts of code exhibited register spills. Therefore, when register spillage was observed, kernels could be fissioned into multiple kernels until register spillage did not occur. This results in larger kernel launch overheads, but significantly lower kernel runtimes. Therefore, there is a balance to strike when using this optimization strategy. Another method to manage kernel launch latencies was via launching all kernels asynchronously in the same stream so that larger kernel runtimes overlap launch overheads for later kernel launches.

To provide application portability across CPUs and NVIDIA, AMD, and Intel GPUs, the Kokkos \cite{9485033} and Yet Another Kernel Launcher (YAKL) \cite{norman2022portable} C++ portability libraries were used. The E3SM-MMF's new Cloud Resolving Model uses C++ Kokkos code for cloud microphysics and macrophysics, and this required interaction between two different C++ portability libraries: Kokkos and YAKL. To accomplish this, Kokkos and YAKL codes exist in separate and self-contained CMake libraries with an interoperation layer provided by YAKL that allows an intermediate representation of multi-dimensional array objects. This way, each C++ portability library can maintain its own CMake build approach, flags, and C++ standard requirements. The YAKL C++ portability library contains a transparent pool allocator for all device-resident allocations so that frequent allocation and deallocation patterns are non-blocking and very cheap, which reduced latencies for device allocations and deallocations.

Finally, part of the ECP funding for E3SM-MMF was devoted to writing a new Cloud Resolving Model, which increases arithmetic intensity via higher-order interpolation and Weighted Essentially Non-Oscillatory (WENO) limiting \cite{norman2020holistic}. This improvement in arithmetic intensity is better suited to GPUs, including AMD GPUs, which have high computational bandwidths compared to data transfer rates. 

\subsection{CoMet}
Another CAAR selection, CoMet~\cite{comet_sc18} is a scientific application used to find similarity correlations between data objects in large datasets, with application to science domains such as genomics, climate, bioenergy and pandemics~\cite{lagergren}.
To do this, CoMet acts on data elements stored as vectors and computes similarity relationships between these vectors, making it possible to identify clusters of items that each represent some common characteristic of scientific interest.
A unique feature of CoMet is its ability to process data using mixed precision arithmetic.
Scientific problems that allow encoding of data as binary values with small numbers of bits, for example in genomics or climatype analysis, are able to take advantage of reduced precision capabilities of GPUs, originally developed to support deep learning workloads.
CoMet can calculate on data using FP32, FP16, Int8 and other datatypes, making it possible to solve much larger problems than would be otherwise possible.

CoMet was designed to use abstractions for its GPU-related internal functions, making it easy to port the code to the HIP programming environment.
To achieve this, the code uses {\tt\#ifdef}s where needed to allow selective compilation for either CUDA or HIP/ROCm backends.
This strategy is also effective managing calls to the required AMD rocBLAS and rocPRIM libraries, whose interfaces are close to or identical to their CUDA counterparts. The CoMet team was able to articulate precise library requirements to AMD early in the project, enabling delivery of high performance routines optimized for the CoMet target problem. On Frontier, CoMet has achieved over 6.71 exaflops of performance using mixed FP16/FP32 arithmetic on 9,074 compute nodes.
Since its computational expense is overwhelmingly dominated by the mixed precision GEMM matrix product operation to compute correlations, CoMet exhibits near-perfect weak scaling behavior up to full system scale.

\subsection{NuCCOR}
NuCCOR (Nuclear Coupled Cluster Oak Ridge) is a DOE Office of Nuclear Physics application developed at ORNL to investigate atomic nuclei from first principles. It solves the time-independent Schr\"odinger equation for many interacting protons and neutrons using the coupled-cluster theory tailored to the symmetries inherent in atomic nuclei. Since its inception in the early 2000s, NuCCOR has migrated from a pure Fortran 95 application to one written primarily in modern Fortran, extensively using object-oriented design patterns to ensure portability between HPC platforms. By writing clean code and creating abstraction layers for all hardware and library dependencies, NuCCOR is robust to many pitfalls that befall HPC (particularly Fortran HPC) applications. For example, adding a new hardware architecture or support for a new library is just a matter of creating the appropriate plugin and adding it to the appropriate factory classes.

Throughout this period, we have also used portability tools like OpenMP and OpenACC to accomplish performance portability. However, solely relying on these tools led to severe roadblocks, mainly because compilers and tools for Fortran applications lag behind their C++ implementations. 

For the Frontier CAAR project, NuCCOR maintained a minimal build where all GPU calls were made with wrappers to C function calls and compiled with a working version of \texttt{gfortran}. Portability is always handled first by abstraction. We added support for new hardware, libraries, and tools in plugins that implement a pre-existing interface without affecting the domain science code. This way CUDA Fortran, \texttt{hipfort}, OpenMP, or any other tool becomes an optional dependency for experimentation instead of a requirement. By writing the scientific code to abstract interfaces instead of actual implementations, we can focus on adding new capabilities instead of constantly porting code to run on new systems.

With this plan in place, the NuCCOR team successfully ported the application to run on Frontier by converting all CUDA code to HIP using the \texttt{hipify} tool and creating the necessary adapters to libraries like rocBLAS. In addition, the C interoperability functionality added to Fortran 2003 provides a straightforward way to use both CUDA and HIP in Fortran without relying on non-standard language extensions only available in selected compilers. Because of this, our work with the COE focused on investigating bugs and performance optimizations that only affected a single plugin. As a result, we made progress on isolated issues without compromising the production build.

\subsection{Pele}\label{pele}

Under the umbrella of the ECP, the Combustion-Pele project has developed applications for performing reactive flow simulations with adaptive mesh refinement (AMR) in complex geometries. Two separate solvers were created: PeleC solves for the fully compressible, multi-species Navier-Stokes equations, while PeleLM(eX) implements a low Mach number approach. Both applications are built upon the AMReX block-structured AMR library~\cite{Zhang:2019} and use embedded boundaries (EB) to represent complex geometries. Both applications share a library called PelePhysics which contains a code generator to emit code for thermo-chemistry routines.

PeleC and PeleLM(eX) began as hybrid C++/Fortran CPU-based applications targeting many-core architectures. To target GPUs, a prototype of PeleC was written in OpenACC as discussed in \cite{PeleC_IJHPCA}. Concurrently, the AMReX library was implementing a performance portability abstraction of its own, using C++ similar to Kokkos~\cite{9485033}, RAJA~\cite{raja}, and GridTools~\cite{AFANASYEV2021100707}. The performance of the OpenACC prototype was found to be equivalent to a similar prototype of PeleC written using the AMReX C++ performance portability library. Although it appeared the Fortran code could persist on the GPU, the Fortran code in PeleC and PeleLM(eX) was abandoned in favor of writing code entirely in C++. One principal reason was that compiler support for Fortran on GPUs was noticeably lacking. Second, it gave the advantage of developing a single code base which could run on CPUs, as well as three different vendor GPU options: NVIDIA, AMD, and Intel. It was also found to be 2x faster on CPUs due to the compiler's ability to optimize a single language. Lastly, the code is easier to debug when using a single language.


Figure~\ref{fig:pelec-time-per-cell-per-timestep} shows an approximate history of the full single node performance timeline of PeleC throughout the project from its many-core beginning to its state on Frontier at the time of writing. Ultimately, a 75x speedup of the code was achieved over the length of the project due to both software and hardware improvements. The computer hardware in the machines shown in Figure~\ref{fig:pelec-time-per-cell-per-timestep} are: NERSC Cori, a many-core Intel Xeon Phi based machine with 68 cores per node; ANL Theta, a many-core Intel Xeon Phi with 64 cores per node; NREL Eagle, an Intel Skylake CPU based machine using two sockets with 36 cores per node; OLCF Summit, an IBM Power9 CPU machine with six NVIDIA V100 GPUs per node; and OLCF Frontier, an AMD EPYC CPU machine with four AMD MI250X Instinct GPUs per node. 



Figure~\ref{fig:pelec-time-per-cell-per-timestep} also shows the performance on 4096 nodes of Summit and Frontier over time.  Since most of the time spent in PeleC and PeleLMeX is in chemistry routines, changes to the chemsitry integrators made over the years  have had an especially notable impact on performance. The initial porting to GPU was the most lucrative increase for single node performance. For AMReX,  the largest performance increase at large scale came from the the asynchronous ghost cell exchange implementation, completed in March of 2021. At the time of writing, weak scaling efficiency of PeleC and PeleLMeX from one to 4096 Frontier nodes is over 80\%.

\input{figures/pele/pelec-time-per-cell-per-timestep}




Optimizations done by the Pele project for on-node performance were generally targeted towards the chemistry routines, while most of the large scale optimization work was done by the AMReX project. Efficiency in the code was gained mostly through the following optimizations: 

\begin{itemize}
    \item The chemistry integration can be performed explicitly or implicitly for non-stiff or stiff chemical mechanisms, respectively. Significant speedup and GPU portability was gained in both solvers by interfacing with the SUNDIALs ODE integrators, in particular CVODE~\cite{Balos:2021}. Instead of our historical point-wise integration of the chemical system, all the cells in the box are assembled into a large chemical system and solved at once with CVODE. In PeleC, a matrix-free GMRES approach is used within the CVODE non-linear solve, minimizing the memory requirements. In PeleLM(eX), batched linear algebra from the MAGMA~\cite{Abdelfattah:2019} library is employed to achieve high throughput and leverage the full potential of CVODE.
    \item Within the chemical integration, the main kernels are the computation of chemical production rates and the chemical Jacobian. These two kernels have been heavily refactored to minimize thread private arrays and explicitly precompute/unroll the code.
    \item Some EB routines require sorting algorithms in which device versions were implemented using existing libraries such as Thrust.
    \item Fused kernel launches were used for larger device throughput when using smaller boxes.
    \item Asynchronous ghost cell exchange were implemented.
\end{itemize}

Porting to Frontier happened through co-design with the AMReX team, as well as other AMReX-based applications such as Nyx~\cite{Nyx}, AMR-Wind~\cite{amr-wind} in the ExaWind project, and WarpX~\cite{10046112}. While AMReX developed and pioneered the interface to the library and the GPU device backends, which compile to each vendor's native programming model, the initial use of unified virtual memory (UVM) allowed each project to adapt their existing code seamlessly. This made it possible to convert the code section by section until full execution on device was achieved. However, removing the use of UVM was ultimately necessary for obtaining better performance on the Frontier AMD platform. With clear similarities between HIP and CUDA, most of the prototyping for the Pele conversion to GPUs was done on Summit, while enabling HIP merely required reporting and resolving bugs. The use of precursor Frontier hardware was also crucial to understanding and accounting for slight differences between the initial NVIDIA hardware and programming models.

Lastly, the Pele applications have been found to stress the machine and compilers in ways that other applications do not. For example, the unrolled chemistry computation routines can contain upwards of 200k lines of code in a single file, with a single GPU kernel (such as the calculation of a chemical Jacobian) spanning 140k lines of code on its own. These large kernels have been found to use upwards of 18k registers. This can result in long compile and link times, and it is apparent that Pele could benefit from further optimization by breaking up these large kernels (i.e., ``kernel fission'').

\subsection{COAST}

The Communication-Optimized All-Pairs Shortest Path (COAST) project
represents a classic graph-theoretical approach to data mining.
The target is mining of scholarly articles, specifically biomedical literature.
The objective is to discover unknown relationships among concepts.
Examples of applications include:
a pharmaceutical laboratory discovering candidate drugs for a disease,
an environmental agency linking a toxin to a medical condition, etc.

The data starts as annotated text and is converted to graph representation
(knowledge graph). The graph commonly comes from the SPOKE database~\cite{spoke}, which integrates data from over 40 sources
into a graph of over 50 million vertices. The vertices represent biological
and biomedical concepts, e.g., genes, diseases, proteins, and symptoms.
The edges represent known relationships among them, such as
``compound causes side effect'' or ``nirmatrelvir/ritonavir treats COVID-19''. 
The code solves the all-pairs shortest path (APSP) problem,
i.e., it finds the shortest path between each pair of nodes in a graph.
It does that using a parallel, distributed, and GPU accelerated version
of the Floyd-Warshall algorithm, which is a canonical example of dynamic programming.

The code was used for a submission to the Gordon Bell competition
in 2020~\cite{kannan2020scalable}. It achieved 136~petaflops
on the Summit supercomputer at ORNL,
and was selected as one of six finalists.
In 2022, the code was ported to the Frontier supercomputer at ORNL
and used for a new Gordon Bell submission.
This time the reported performance achieved \mbox{1.004~exaflops}, and the submission
was selected as a finalist again~\cite{kannan2022exaflops}.
The greater than 7x performance increase resulting from porting from NVIDIA hardware to AMD hardware was achieved with a few key strategies:

\begin{itemize}
    \item For portability from the CUDA API to the HIP API,
    the code relies on a thin layer of abstraction that defines functions
    like \verb|set_device()| and \verb|device_stream_create()|, and
    delegates execution to \verb|cudaSetDevice()| and \verb|cudaStreamCreate()|
    or \verb|hipSetDevice()| and \verb|hipStreamCreate()|, depending on the compile-time
    configuration.
    \item For achieving high performance on AMD GPUs, the code relies on automated software tuning. 
    For example, the main computational kernel, which heavily resembles matrix multiplication, is written in C++ as nested loops with multiple levels of tiling, and the best set of tiling factors is discovered in the process of compiling and timing a large number of combinations.
\end{itemize}

The kernel optimization process was successful, as the performance increased from \mbox{5.6~teraflops} on one NVIDIA Volta GPU of the Summit supercomputer to \mbox{30.6~teraflops} on one AMD Instinct\texttrademark
 MI250X GPU of the Frontier supercomputer.

\subsection{LAMMPS}\label{LAMMPS}
LAMMPS~\cite{LAMMPS} is a classical molecular-dynamics simulator code focusing on materials modeling on parallel computers.
LAMMPS has multiple different back-ends, delivering highly-optimized performance for a number of different execution models (e.g., OpenMP~\cite{OpenMP98}, OpenCL~\cite{OpenCL10}, and other vendor-specific options) for a wide variety of force-field models, constraints and simulation configurations.
On Frontier, the focus was on performance attainment of the Kokkos~\cite{9485033,CarterEdwards20143202} back-end of LAMMPS targeting HIP~\cite{OLCFHIP,AMDHIP} for the ReaxFF~\cite{doi:10.1021/jp709896w,AKTULGA2012245,doi:10.1021/jp201599t} force-field simulation of crystalline Hexanitrostilbene (HNS).

This work comprised several parts:
\begin{itemize}
    \item Identification (and fixing) of significant correctness issues in AMD's Clang~\cite{LLVM:CGO04}-based compiler for HIP.
    \item Algorithmic co-design and optimization of LAMMPS' implementation of ReaxFF and Kokkos' HIP back-end.
    \item Optimization of AMD libraries and compilers.
\end{itemize}

\subsubsection{Fixing correctness issues}
The development of a HIP backend for Kokkos was achieved via a collaborative effort with several members of the Kokkos team at the Scalable Algorithms and Coupled Physics Group at OLCF. After enabling functionality, efforts shifted to running and optimizing LAMMPS.
While initial tests, particularly for simpler force-field styles (e.g., a Lennard-Jones potential) ran without significant issues, the more computationally intensive ReaxFF model initially caused intermittent segmentation faults and out-of-bounds memory accesses at runtime.

This issue proved to be complicated to solve, as small changes to the compiler (code-generation), source-code, or even run-to-run timing variations could be enough to trigger these types of faults intermittently at any number of source-code lines.
Ultimately, obtaining correctness required extensive, close, and direct collaboration between application teams and compiler experts at AMD.
The issue was ultimately tracked to \href{https://reviews.llvm.org/D124195}{register spills in highly divergent code-regions}.
Kokkos' portability proved to be a significant help in the development of the compiler fix: by relaxing some of the more strict memory access requirements (e.g., to enable direct GPU-memory access from the CPU over LargeBAR), the same ``kernel'' could be run on both the CPU and GPU, using the same memory allocations (at a significant latency penalty on the CPU).  
This enabled fine-grained correctness validation of the calculated forces and other intermediate values.

\subsubsection{Algorithmic co-design and optimization}

Once the Kokkos-HIP back-end of LAMMPS was reliably running ReaxFF, the focus shifted towards performance attainment.
Initial profiling on AMD Instinct GPUs found a few key bottlenecks:
\begin{itemize}
    \item High thread-divergence in the ReaxFF force-field evaluation kernels.
    \item Low occupancy in some ReaxFF force-field evaluation kernels due to register-pressure constraints.
    \item Memory bandwidth requirements for the sparse matrix-vector multiplications inside of the partial charge equilibration routines, and related communication overhead.
\end{itemize}

\begin{algorithm}
\caption{Pseudo-code for Torsion-force evaluation in ReaxFF}\label{alg:reaxff_torsion}
\begin{algorithmic}
\Procedure{eval\_torsion}{$i,neigh*,bond*$}
\For{$j$ \textbf{in} $neigh[i]$}
    \If{\emph{cutoff}$(i, j)$}
        \For{$k$ \textbf{in}  $neigh[i]$}
            \If{\emph{cutoff}$(i, k)$}
                \For{$l$ \textbf{in} $bond[j]$}
                    \If{\emph{cutoff}$(i,j,k,l)$}
                        \emph{torsion}$(i,j,k,l)$
                    \EndIf
                \EndFor
            \EndIf
        \EndFor
    \EndIf
\EndFor
\EndProcedure
\end{algorithmic}
\end{algorithm}

A highly-idealized version of the original code-pattern for evaluation of the Torsional force component of the ReaxFF model is presented in Algorithm~\ref{alg:reaxff_torsion} to help visualize the sources of divergence.

The outer atom index $i$ is initialized to the global HIP thread index, and marched as a grid across all atoms in the system.
In addition, the procedure is passed two lists, a distance-based neighbor-list (\emph{neigh}) and an atomic bond based neighbor-list (\emph{bond}).
The \emph{cutoff} function takes into account the distance (and bonds) between two, or four, atomic coordinates and returns \emph{True} if the pair-wise force should be evaluated.
Finally, if a tuple of $(i, j, k, l)$ indices makes it through all of the cutoff checks, the \emph{torsion} function is called, containing many expensive memory loads and floating-point operations.

This pattern appeared in the evaluation of Angular and Torsional force-field terms in ReaxFF.
In the worst case (the Torsion example above), profiling indicated that on average only a handful of threads in the entire wavefront were active.
Optimization of this pattern depended on a key observation: the cost of evaluating the \emph{cutoff} function is proportionally small as compared to evaluation of the full \emph{torsion} (or other) force.
Through the \href{https://github.com/lammps/lammps/pull/3147}{efforts} of multiple \href{https://github.com/lammps/lammps/pull/3195}{collaborators}, a significantly faster method was developed: a ``preprocessor'' kernel is launched that computes a list of successful $(i, j, k, l)$ interaction tuples.
Then, the Angular and Torsional force-field kernels consume this precomputed list, such that almost \textit{all} of the control flow in Alg.~\ref{alg:reaxff_torsion} can be eliminated, and the pairwise force terms can be evaluated in a ``dense'' manner.
This methodology was also \href{https://github.com/lammps/lammps/pull/3158}{extended} to the kernel responsible for creating the \emph{bond} neighbor-lists themselves for further speedups.

In addition, it was noted that one key optimization made by Aktulga et al.~\cite{AKTULGA2012245} was not present in the Kokkos-backend's implementation of ReaxFF of LAMMPS (it was available in the OpenMP\textsuperscript{\tiny\textregistered} backend, however), namely the joint iteration of two conjugate gradient (CG) loops in the partial-charge equilibration phase of the computation. 
As this computation involves a sparse-matrix vector product operation, wherein the same matrix is multiplied by (and accumulated to) two separate vectors, significant reductions in bandwidth requirements can be achieved by fusing the CG solve loops.
In addition, this scheme greatly reduces the overall number of CG solve loop iterations required for both solutions to converge.
As each CG-solve loop iteration has a significant communication phase (that scales poorly~\cite{AKTULGA2012245}), this also results in a large reduction in the overall time spent in communications during the partial-charge equilibration phase.

The result of this joint-effort was a greater than $50\%$ speedup of ReaxFF in LAMMPS since Feb. 2022 for multiple GPU-vendors~\cite{copa2022}: a success for both application performance and portability.

\subsubsection{Optimization of AMD libraries and compilers}

The final thrust of optimization effort in readying LAMMPS for Exascale was through improving the compiler and device-libraries shipped as part of the ROCm stack.
During intial performance work, it was noted that a handful of the key kernels \textemdash\ in particular, the Torsional and Angular pair-wise force evaluations discussed previously\textemdash\ had a significant number of vector register spills, as determined by the \textit{vgpr\_spill\_count} and \textit{amdhsa\_private\_segment\_fixed\_size} fields in the assembly dumps (from the output of the compiler's \textit{--save-temps} option).
Using DWARF~\cite{dwarf} information, and working with AMD Compiler engineers, this issue was tracked back to inefficiencies in \href{https://reviews.llvm.org/D104874}{spilling of double-precision constants between scalar and vector registers}.
Along with some changes to the register allocation scheme, this virtually eliminated register spills from the key kernels.
Finally, microbenchmarking the achieved throughput of some heavily used math functions (e.g., \textit{pow()} and \textit{exp()}) exposed some \href{https://github.com/RadeonOpenCompute/ROCm-Device-Libs/commit/7f732ad2d2b57bff90a5623d44e2e8aefc9138a3}{additional optimization opportunities}.

\section{Access to Early Hardware Platforms} \label{early-hw}

A key enabler for application readiness was the provision of early systems with successive generations of hardware well before the final exascale system became available. The purpose of these early access systems was to allow users to familiarize themselves with the HPE Cray Programming Environment and AMD software stack, aid in the porting of applications to HPE Cray systems,
and provide a platform for the tuning of applications for AMD GPUs. 

From the inception of the project, HPE, AMD, and OLCF worked together to ensure that key application teams had access to early platforms as early as possible. In addition, the platforms were constructed to give application teams a development environment that would ultimately converge on the target exascale platform. Whenever possible, these systems were accompanied by talks and documentation (see Section~\ref{training}) detailing how the accessible platform differed from the final system node architecture. This ensured application teams wasted as little time as possible in optimizations that would not benefit the final target exascale platform. 

Starting in 2019, three generations of early access platforms were deployed. These systems spanned three generations of AMD Instinct GPU hardware (MI60s, MI100s, and the MI250X products used in Frontier), multiple generations of the HPE Slingshot Interconnect, and three generations of AMD EPYC CPUs. 
Access to the systems was restricted to teams that were under non-disclosure agreements (NDAs) permitting sharing information on the future hardware and system configurations. Users on the systems were permitted to publish results after the hardware in the clusters was made generally available.

In general, all three generations of early access hardware platforms were useful development platforms. All the platforms shared key commonalities, in that they were heterogeneous systems, based on AMD compute hardware, and shared a common instruction set (x86 on the EPYC CPU and AMD's Compute DNA architecture for the Instinct GPUs). All early access systems had software and tools that would be available on Frontier, in particular the HPE Cray Programming Environment and AMD's ROCm software stack. 

The first generation platforms were named ``Poplar'' and ``Tulip'' and were based on AMD MI60s and AMD EPYC 7601 ``Naples'' CPUs. The second generation systems (named ``Spock'' and ``Birch'') had AMD Instinct MI100s and AMD EPYC 7662 ``Rome'' CPUs. 
They also had the HPE Slingshot interconnect with a 100 GbE interface network. 
Spock and Birch had six nodes and 12 nodes, respectively, each with four GPUs in each node. 
These systems were much closer in software stack and interconnect to the eventual Frontier system, and were of sufficient scale to permit modest scaling studies.
Developers were able to access both systems in 2020, which was considerably before access to Frontier. 

The final system (named ``Crusher'') is identical to the Frontier node architecture, with AMD 64-core Optimized 3rd gen EPYC CPUs and AMD Instinct MI250X GPUs. The system also has the HPE Slingshot interconnect with 200 Gigabit Ethernet (GbE) interfaces. Crusher was delivered well in advance of the general system availability, permitting early access users to begin tuning on the system in January, 2022. With 192 compute nodes, each containing 4$\times$ MI250X, the system was capable of providing early experiences to users scaling applications to hundreds of compute devices on software close to the stack used on Frontier, the production system.

A key strategy for leveraging the early systems was to test application use
cases that precisely mimic the per-GPU problem dimensions expected on exascale
runs. Math libraries achieve maximum performance through tuning for the
complex hierarchy of memory levels and device parallelism of GPUs. Performance
trade-offs depend on specifics of the input and output sizes, so libraries
often contain a large collection of problem-size-dependent implementations.
Early access allowed application developers to provide target problem sizes for library developers, such that the libraries were tuned and ready for these applications when the final system arrived.






\section{Hackathons, Trainings, User Guides} \label{training}

Select application and software teams (henceforth “early users”) had extensive access to AMD, HPE and OLCF staff for expert consulting and knowledge sharing. Early users were given access to the early-access hardware platforms described in Section ~\ref{early-hw}. The OLCF, in coordination with HPE and AMD, created a quick-start guide and organized a training workshop for each system to provide a baseline for system access and application porting. From there, any questions or issues encountered by the users were addressed through OLCF support tickets (and a COE Confluence\texttrademark site). Each application team was paired with liaisons affiliated with the COE who could provide expert knowledge and flexible support for high-priority CAAR and ECP projects. The liaisons participated in regular virtual meetings, instant-messaging channels, wikis, software repositories, issue-tracking systems, and direct E-mail. Many of the strategies documented in Sections~\ref{software} and \ref{applications} were the result of close interactions between liaisons and application teams. Interactions were disrupted in part by the COVID-19 pandemic in 2020. Outside of the CAAR Kick-Off workshop the week of Oct 7th 2019, nearly all the interactions were virtual. 

Shortly after the early users obtained access to an early access system, such as Spock or Crusher, the OLCF, HPE, AMD, and ECP would hold a ``hackathon'' for subsets of the early user teams. The idea was to give the teams access to the new system, let them attempt to get their codes up and running, then follow up with hackathons to help the teams drive toward their goals (e.g., compiling, optimizing, and debugging). These hackathons proved to be a great source for identifying common user issues and software bugs, and generally collecting lessons learned. The lessons learned from the hackathons were then disseminated to the rest of the early users (and facility and vendor staff) through special webinar sessions. 
Then the information was further distilled into new sections in the user guide to help all (current and future) users of the system\footnote{see for example \url{https://docs.olcf.ornl.gov/systems/crusher_quick_start_guide.html}}.

Trainings covered a wide spectrum of topics across hardware, software and system operations. Topics in hardware included information pertinent to optimization, such as cache sizes, hardware atomics, register spilling, and kernel launch latencies. Software topics included code examples to leverage new features, such as specialized SGEMM/DGEMM operations, the AMD Infinity Fabric Interconnect, or HIPifying codes. System topics included call patterns for the batch system and NUMA and affinity considerations. 

\section{Lessons Learned and Conclusions}
\label{lessonslearned}

The most important lesson learned from the Frontier experience was that application readiness on leadership computing resources does not occur without deliberate support across a spectrum of modalities including: hardware systems, CPU and GPU software, performance tools/debuggers, networking software, and application performance tuning and optimization expertise. The close collaboration inherent to the COE across HPE, AMD, and ORNL ensured a thoughtful allocation of resources and effective knowledge sharing between code teams, vendor staff, and domain experts. 

As detailed in Sections \ref{software} and \ref{applications}, enabling applications to run on Frontier was most successful when accelerated codebases existed, where porting and optimization to AMD's HIP was an efficient approach that also provided portability between multiple GPU vendors. Applications also had success running on Frontier when using abstraction frameworks such as Kokkos/RAJA, or OpenMP target offload directives. Best performance was achieved when applications leveraged vendor provided libraries, and when the application teams were able to set clear performance targets and provide precise function dependencies and problem sizes early in the project. 

Many of the applications in Section~\ref{applications} achieved significant speed-ups, with performance improvements between 5x and 7x vs. OLCF Summit (on a per device or scaled-out basis) being typical. Several applications also demonstrated sustained performance above an exaflop on real scientific problems. 
Overall, the success of these projects indicates that exascale computing is a paradigm shift in capability, not in programming models or optimization techniques. Traditional techniques such as HIP/CUDA, OpenMP, and MPI are still valid to achieve high performance on modern supercomputers. In particular, the ``GPU-Aware MPI + X'' model for inter-node communication remains the predominant narrative for Frontier and the exascale era.

Application readiness was also accelerated by early hardware and software access. Early access to software and hardware helped identify:  A) functionality problems, B) missing features, and C) performance problems, typically in this order. Platforms were seldom ``too early'' for initial assessments and testing. Rather, software issues that were identified earlier provided more time for vendors to identify and implement a fix. The communication of software issues and their fixes was also an important activity. Documenting known performance issues, and their mitigation, on Confluence pages or during virtual meetings saved COE early-access users considerable time while tuning and porting codes and avoided multiple teams triaging the same issue.

The COE also greatly benefited from a quantitative approach to tracking application readiness. Application teams were expected to provide a well-posed challenge problem and figure of merit (FOM) on Summit and an acceleration plan for Frontier. The teams then produced  mid-project reports (reviewed by the COE Management Council) and a final report detailing challenge problem results. This quantitative approach permitted early detection of software bugs and performance regressions, and enabled continuous assessment of applications against their stated speed-up targets.

If this project were done again, the COE would do several things differently. Clear information would be provided to users on what particular version of the CUDA API was supported by AMD's HIP. Similarly, the COE would communicate what CUDA functions and capabilities were not likely to be supported in HIP (likely due to hardware differences). Finally, more training on common issues (atomic operations, CPU and GPU bindings, and compiler optimizations )would be provided as early as possible to users. 

The intensive engagement among the compute vendors, OLCF, and application teams through the COE has proved invaluable in preparing applications for exascale. It is expected that the function of the COE will remain vital for future leadership computing systems, which are likely to experiment with advanced computer architectures and programming models. There are several examples of the efficacy of machine learning and artificial intelligence for scientific computing~\cite{9563024,McDougall2017,10.1145/3392717.3392772}, and so it is also likely that the role of the COE will evolve as applications evolve to leverage these approaches. 
\begin{acks}
This research was supported by the Exascale Computing Project (17-SC-20-SC), a collaborative effort of the DOE Office of Science (SC) and the NNSA, and was performed using computational resources of the Oak Ridge Leadership Computing Facility, which is a DOE SC User Facility supported under Contract DE-AC05-00OR22725

This manuscript has been co-authored by Oak Ridge National Laboratory, operated by UT-Battelle, LLC under Contract No. DE-AC05-00OR22725 with the U.S.Department of Energy. 
The views expressed in the article do not necessarily represent the views of the DOE or the U.S. Government. The U.S. Government retains and the publisher, by accepting the article for publication, acknowledges that the U.S. Government retains a nonexclusive, paid-up, irrevocable, worldwide license to publish or reproduce the published form of this work, or allow others to do so, for U.S. Government purposes. DOE will provide public access to these results of federally sponsored research in accordance with the DOE Public Access Plan.

AMD, the AMD Arrow logo, Instinct, EPYC, and combinations thereof are trademarks of Advanced Micro Devices, Inc. Other product names used in this publication are for identification purposes only and may be trademarks of their respective companies.
\end{acks}

\printbibliography

\appendix




\end{document}

%% file: figures/pele/pelec-time-per-cell-per-timestep.tex
\begin{figure}
\centering
\begin{tikzpicture}

\begin{groupplot}[group style={group size= 1 by 2}, height=8cm, width=10cm]
\nextgroupplot[
  xbar, xmode=log, log origin=infty, bar width=4pt,
  enlarge y limits=0.2,
  title=PeleC Time Per Cell Per Timestep,
  yticklabel style={font=\tiny},
  xlabel={Normalized Time},
  xmin=0, 
  ytick=data,
  symbolic y coords={Mar 2023 Frontier, Nov 2021 Perlmutter, Nov 2021 Summit, Sep 2020 Summit, Sep 2020 Eagle, Sep 2020 Theta, Sep 2018 Cori},
  legend pos=south east,
  legend columns=1,
  legend style={font=\tiny},
  width=0.8\linewidth, height=0.4\linewidth,
  every axis plot/.append style={thick},
]

\addplot coordinates {
(1.000000000,Sep 2018 Cori)
(2.386770628,Sep 2020 Theta)
(1.387950205,Sep 2020 Eagle)
(0.059273098,Sep 2020 Summit)
(0.026246621,Nov 2021 Summit)
(0.019587354,Nov 2021 Perlmutter)
(0.013133049,Mar 2023 Frontier)
};

\legend{1 Node}

\nextgroupplot[
  xbar, bar width=8pt,
  enlarge y limits=0.2,
  yticklabel style={font=\tiny},
  xmin=0, 
  ytick=data,
  symbolic y coords={Mar 2023 Frontier, Nov 2021 Summit, Sep 2020 Summit},
  legend pos=south east,
  legend columns=1,
  legend style={font=\tiny},
  width=0.8\linewidth, height=0.4\linewidth,
  every axis plot/.append style={thick},
]
\pgfplotsset{cycle list shift=1}

\addplot coordinates {
(1.0,Sep 2020 Summit)
(0.201429658301887,Nov 2021 Summit)
(0.08378027074873179,Mar 2023 Frontier)
};

\legend{4096 Nodes}
\end{groupplot}
\end{tikzpicture}
\caption{History of PeleC time per cell per timestep for a single node between September 2018 and March 2023 on a variety of machines. Also shown is the time reduction at the scale of 4096 nodes for the 2020, 2021, and 2023 code states on Summit and Frontier.}
\label{fig:pelec-time-per-cell-per-timestep}
\end{figure}
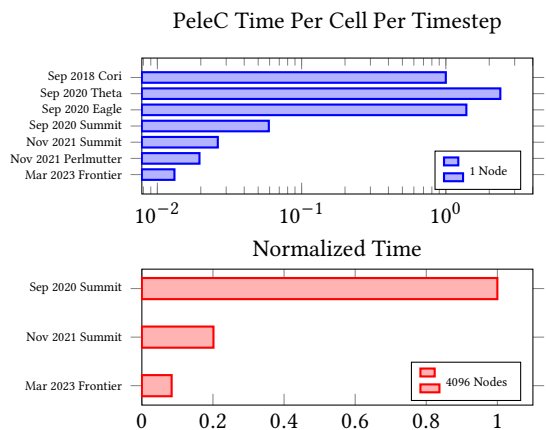